\documentclass[aps,prd,superscriptaddress,nofootinbib,tighten,preprint]{revtex4}

\newcommand{\be}{\begin{eqnarray}}
\newcommand{\ee}{\end{eqnarray}}

\newcommand{\ms}{\Delta m^2_{21}}
\newcommand{\ma}{\Delta m^2_{31}}

\def\gtap{\ \raisebox{-.4ex}{\rlap{$\sim$}} \raisebox{.4ex}{$>$}\ }

\def\gs{\mathrel{
   \rlap{\raise 0.511ex \hbox{$>$}}{\lower 0.511ex \hbox{$\sim$}}}}
\def\ls{\mathrel{
   \rlap{\raise 0.511ex \hbox{$<$}}{\lower 0.511ex \hbox{$\sim$}}}}
\newcommand{\bea}{\begin{equation} \begin{array}{c}}
\newcommand{\bead}{\begin{equation} \begin{array}{cccc}}
\newcommand{\eea}{ \end{array} \end{equation}}
\usepackage[utf8]{inputenc}
\usepackage{amsfonts}
\usepackage{amssymb}
\usepackage{amsmath}
\usepackage{graphicx}
\usepackage{hyperref}
\usepackage{placeins}
\usepackage{gensymb}
\usepackage[T1]{fontenc}
\usepackage{color}
\def\slc#1{\setbox0=\hbox{$#1$}           
    \dimen0=\wd0                                 
    \setbox1=\hbox{/} \dimen1=\wd1               
    \ifdim\dimen0>\dimen1                        
       \rlap{\hbox to \dimen0{\hfil/\hfil}}      
       #1                                        
    \else                                        
       \rlap{\hbox to \dimen1{\hfil$#1$\hfil}}   
       /                                         
    \fi}

\newcommand{\beq}{\begin{equation}}
\newcommand{\eeq}{\end{equation}}
\newcommand{\beqa}{\begin{eqnarray}}
\newcommand{\eeqa}{\end{eqnarray}}

\newcommand{\tx}{{\theta_{12}}}
\newcommand{\ty}{{\theta_{13}}}
\newcommand{\tz}{{\theta_{23}}}

\newcommand{\da}{\delta_{13}}
\newcommand{\db}{\delta_{24}}

\newcommand{\tmet}{\theta^{3\nu}_{\mu e}}
\newcommand{\tmef}{\theta^{4\nu}_{\mu e}}

\begin{document}

\title{Exploring Fake Solutions in the Sterile Neutrino Sector at Long-Baseline Experiments}

\author{Sandhya Choubey}
\email{sandhya@hri.res.in}
\affiliation{Harish-Chandra Research Institute, HBNI, Chhatnag Road, Jhunsi, Allahabad 211 019, India}
\affiliation{Department of Physics, School of
Engineering Sciences, KTH Royal Institute of Technology, AlbaNova
University Center, 106 91 Stockholm, Sweden}

\author{Debajyoti Dutta}
\email{debajyoti.dutta@dbuniversity.ac.in}
\affiliation{Assam Don Bosco University, Tapesia Campus, Sonapur, Assam, 782402 India}

\author{Dipyaman Pramanik}
\email{dipyamanpramanik@hri.res.in}
\affiliation{Harish-Chandra Research Institute, HBNI, Chhatnag Road, Jhunsi, Allahabad 211 019, India}

\begin{abstract}
Active-sterile neutrino mixing is known to affect the neutrino oscillation probabilities at both short as well as long-baselines. In particular, constraints on active-sterile neutrino oscillation parameters can be obtained from long-baseline experiments such as T2HK and DUNE. We present here existence of fake solution in the appearance channel for the 3+1 scenario at long-baseline experiments. We show that the appearance probability is same for values of $\Delta m_{41}^2$ for which the fast oscillations are averaged out and for $\Delta m_{41}^2=(1/2)\Delta m_{31}^2$. The fake solution does not appear for the disappearance channel. 
\end{abstract}
\maketitle      

\section{Introduction}
Neutrino oscillation physics has entered the precision era. Now we know beyond all doubts that there are at least two massive neutrino mass eigenstates and three non-zero mixing angles, $\theta_{12}$, $\theta_{13}$ and $\theta_{23}$. We also have achieved reasonable precision in the measurement of the two mass squared difference and the mixing angles $\theta_{12}$ and $\theta_{13}$. While we have measured $\theta_{23}$, some work remains to be done in this sector. In particular, although there are hints from recent experiments like NOvA \citep{Adamson:2017gxd} and T2K \citep{Abe:2018wpn} that  $\theta_{23}$ is close to maximal, it is still ambiguous. Also, if it was to be non-maximal, we do not know the octant of $\theta_{23}$, {\it .i.e.}, whether it is less than $45^{\circ}$ or greater than $45^\circ$. While there are hints of $\delta_{CP}=-90^{\circ}$ from the current experiments, we still need to confirm that there is CP-violation in the leptonic sector. Finally, we still need to determine the sign of $\Delta m_{31}^2$, {\it aka}, the neutrino mass hierarchy.  Dedicated future experiments such as INO \citep{Kumar:2017sdq}, JUNO \citep{An:2015jdp,Djurcic:2015vqa}, DUNE \cite{Acciarri:2016ooe,Strait:2016mof,Acciarri:2015uup,Acciarri:2016crz} and T2HK \cite{Abe:2015zbg,Abe:2016ero} are being planned to resolve these issues. 


Another unsettled issue is the question regarding the existence of a light sterile neutrino mixed with the active ones \cite{Abazajian:2012ys}. Hints of this came when LSND \cite{Athanassopoulos:1995iw,Aguilar:2001ty} in USA reported $3.8 \sigma$ excess in their positron events, pointing at $\bar\nu_\mu \to \bar\nu_e$ oscillations. This excess could not be explained within the standard three-generation paradigm because the LSND $L/E$ demanded a $\Delta m^2 \sim $ eV$^2$ while, the solar and atmospheric neutrino data demand $\Delta m^2 \sim 10^{-5}$ eV$^2$ and $10^{-3}$ eV$^2$, respectively. Therefore, one has to postulate the existence of a fourth light neutrino such that it corresponds to a mass-squared difference $\Delta m^{2}\sim 1$ eV$^2$. Since the bound on number of light neutrinos coupled to the $Z$-boson is $2.9840\pm0.0082$ from LEP \citep{ALEPH:2005ab}, the fourth generation must be {\it sterile}. Existence of these oscillations were tested at the short-baseline experiments KARMEN \cite{Gemmeke:1990ix} and MiniBooNE \cite{AguilarArevalo:2007it,Aguilar-Arevalo:2013pmq,AguilarArevalo:2010wv} which looked for $\bar\nu_\mu \to \bar\nu_e$ appearance, as well as at CDHS \citep{Dydak:1983zq},  MINOS \citep{Adamson:2010wi}, MiniBooNE (disappearance search) \citep{Cheng:2012yy} and SuperKamiokande atmospheric experiment \citep{Abe:2014gda} which looked at $\nu_\mu(\bar\nu_\mu)$ disappearance. Data from neutrinos with longer baseline such as MINOS \citep{MINOS:2016viw}, IceCube \citep{TheIceCube:2016oqi} and NOvA \citep{Adamson:2017zcg} have also reported null signal for $\nu_\mu\to \nu_e$ oscillations at the LSND scale. 
While KARMEN did not find any evidence for neutrino oscillations consistent with LSND, it was unable to eliminate the entire parameter space favored by LSND. None of the disappearance searches observed any evidence for neutrino oscillations consistent with LSND, making the LSND signal anomalous, which reflects in a very low goodness-of-fit reported from global analyses of all experiments \citep{Gariazzo:2017fdh}.  On the other hand, the MiniBooNE experiment designed specifically to test the LSND anomaly, ran in both neutrino and antineutrino mode and the collaboration have recently presented the complete analysis of their entire data set \citep{Aguilar-Arevalo:2018gpe}. They reported an observed $4.8 \sigma$ excess of electron-like events at MiniBooNE. When combined with LSND this constitutes a $6.1 \sigma$ excess, putting the ``LSND anomaly'' at a statistically compelling level. 

Hints for active-sterile neutrino oscillations have also come from the $\nu_e$ and $\bar\nu_e$ sectors. While the Gallium experiments \citep{Abdurashitov:2005tb,Laveder:2007zz,Giunti:2006bj,Giunti:2010zu,Giunti:2012tn} require oscillations of $\nu_e$ with $\Delta m^2 \sim 1$ eV$^2$, there appear to be inconsistency between the reactor flux calculations and reactor data \citep{Mention:2011rk} which favors existence of a fourth neutrino. Confirmation for active-sterile mixing driven $\bar\nu_e$ disappearance searches have been performed by NEOS \citep{Ko:2016owz}, DANSS \citep{Alekseev:2018efk} and STEREO \citep{Almazan:2018wln} and so-far there is no convincing evidence for active-sterile mixing.


The impact of active-sterile mixing in long-baseline experiments have been studied extensively in the literature. The effect of active-sterile mixing on measurement of standard three-generation neutrino oscillation parameters at long-baseline experiments NOvA, T2HK and DUNE was studied in detail in \cite{Gandhi:2015xza,Dutta:2016glq,Agarwalla:2016xxa,Choubey:2017cba,Gupta:2018qsv,Ghosh:2017atj,Agarwalla:2016mrc,Agarwalla:2016xlg}. On the other hand, possibility of measuring the active-sterile mixing parameters was explored extensively for DUNE near detector in \cite{Choubey:2016fpi} and NOvA, T2HK and DUNE far detectors in \cite{Berryman:2015nua,Kelly:2017kch,Choubey:2017ppj,Gandhi:2017vzo,Coloma:2017ptb}. In this paper we point out a new fake solution that exists in the appearance channel at long-baseline experiments. We will show both analytically as well numerically that the appearance probability is the same for values of $\Delta m_{41}^2$ for which the fast oscillations are averaged out and for $\Delta m_{41}^2=(1/2)\Delta m_{31}^2$. The fake solution does not appear for the disappearance channel. 

The paper is organised as follows: In section~\ref{sec:mixangle} we will show the existence of the fake solution analytically and numerically.  In section~\ref{sec:ext} we will present the main $\chi^2$ results of the paper. Finally, we will conclude in section~\ref{sec:conclusion}.

\section{Fake Solution in the Appearance Probability  \label{sec:mixangle}}

\begin{figure}
\includegraphics[width=0.45\textwidth]{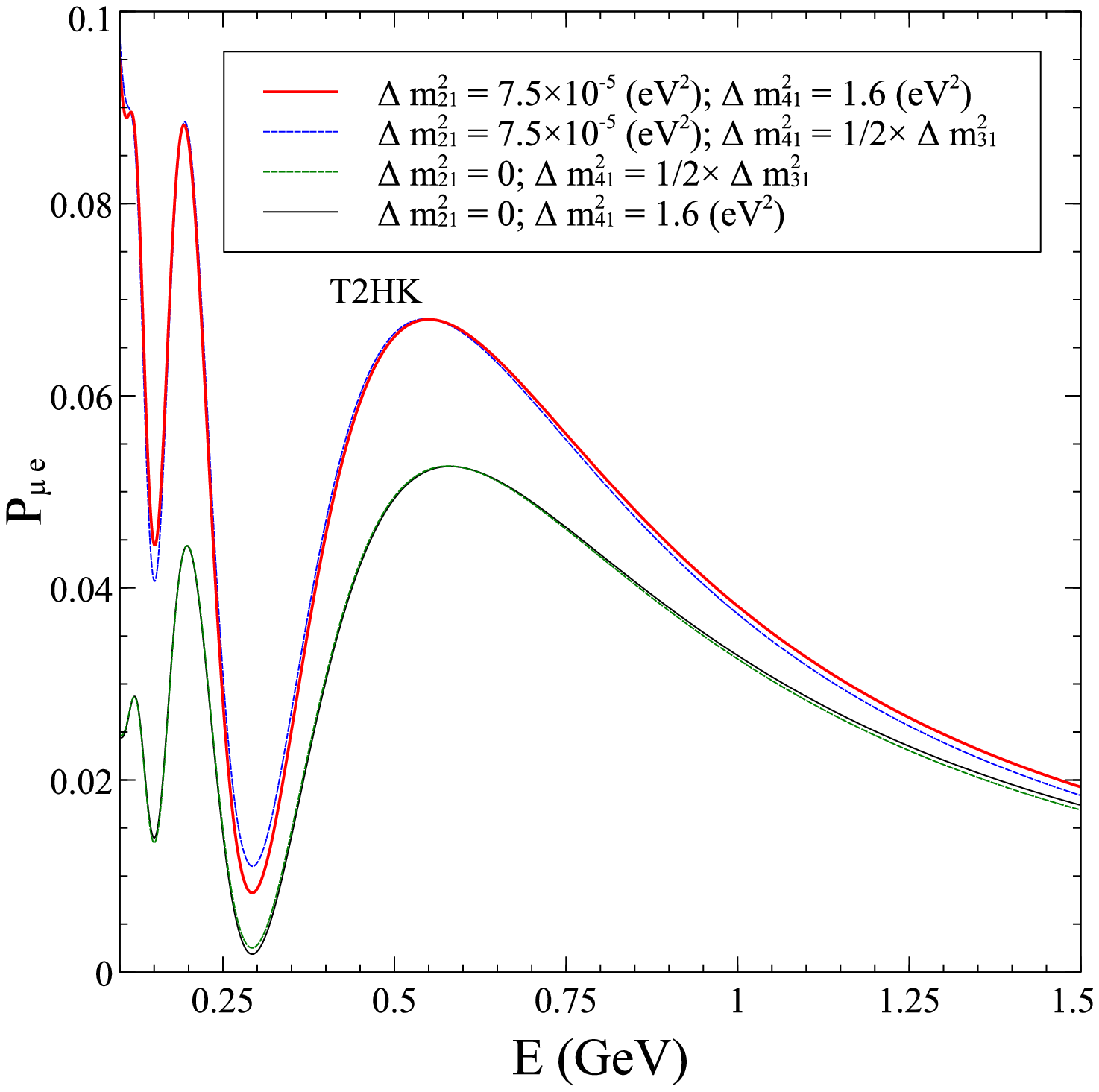}
\includegraphics[width=0.45\textwidth]{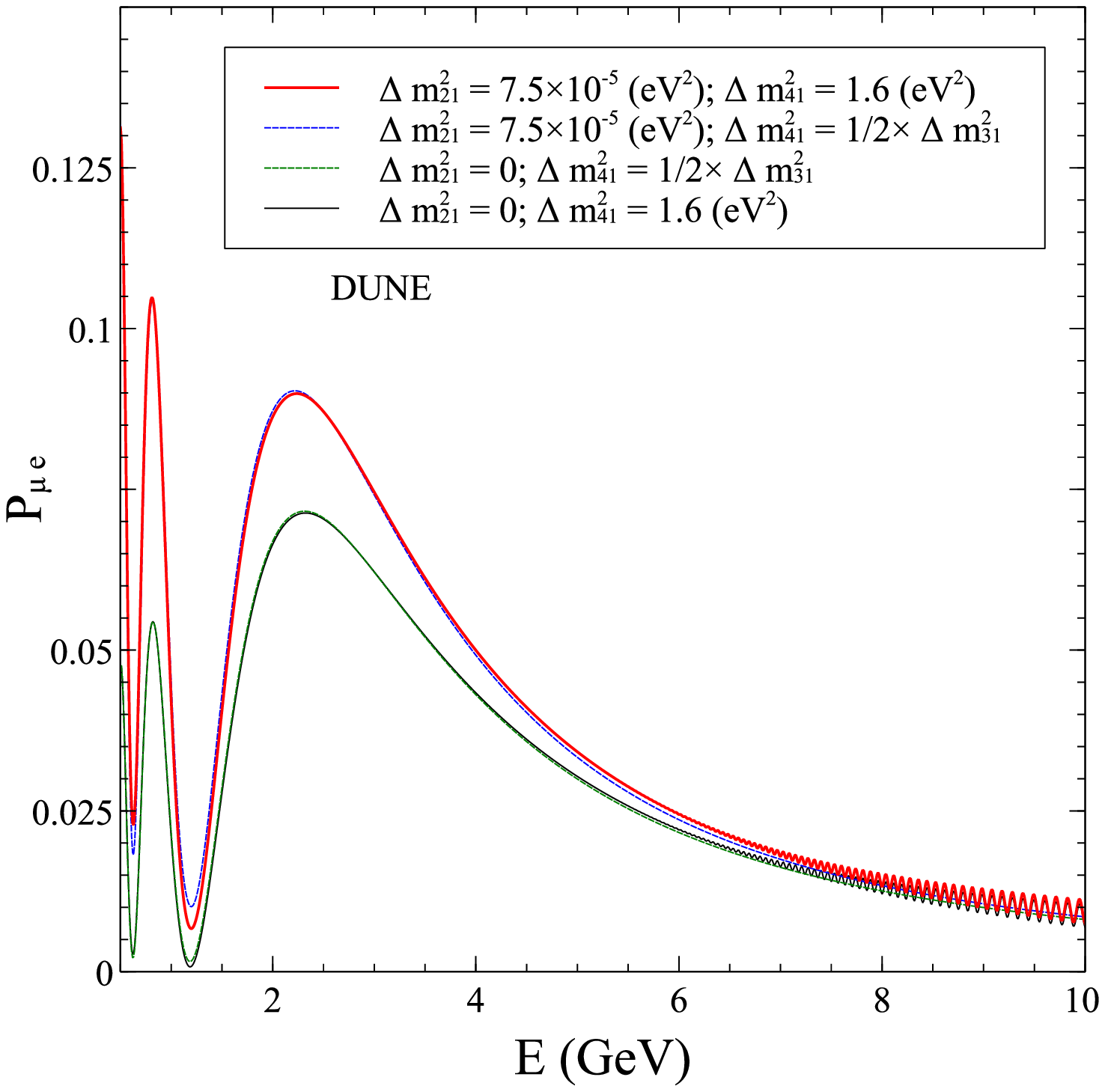}
\includegraphics[width=0.45\textwidth]{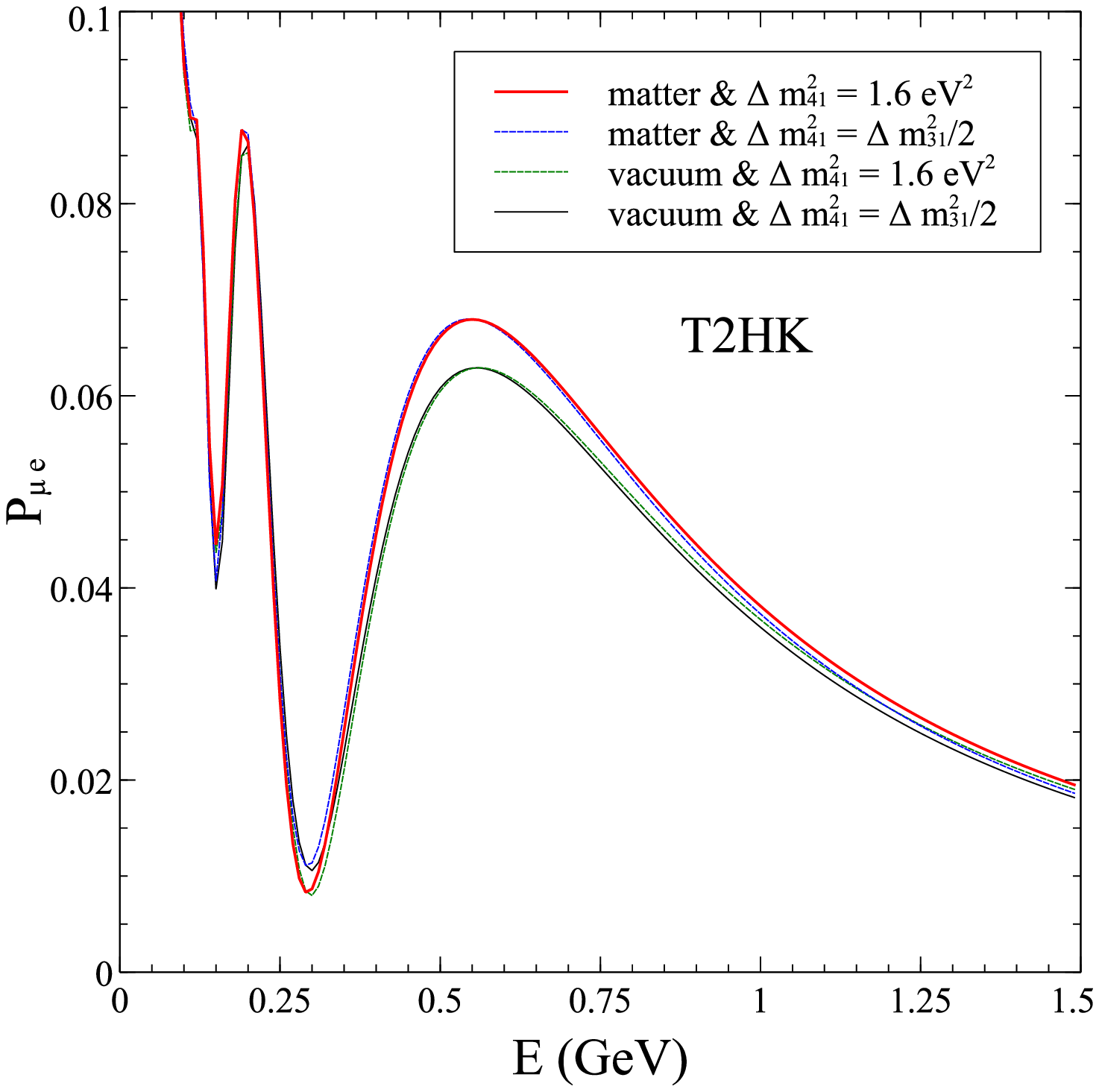}
\includegraphics[width=0.45\textwidth]{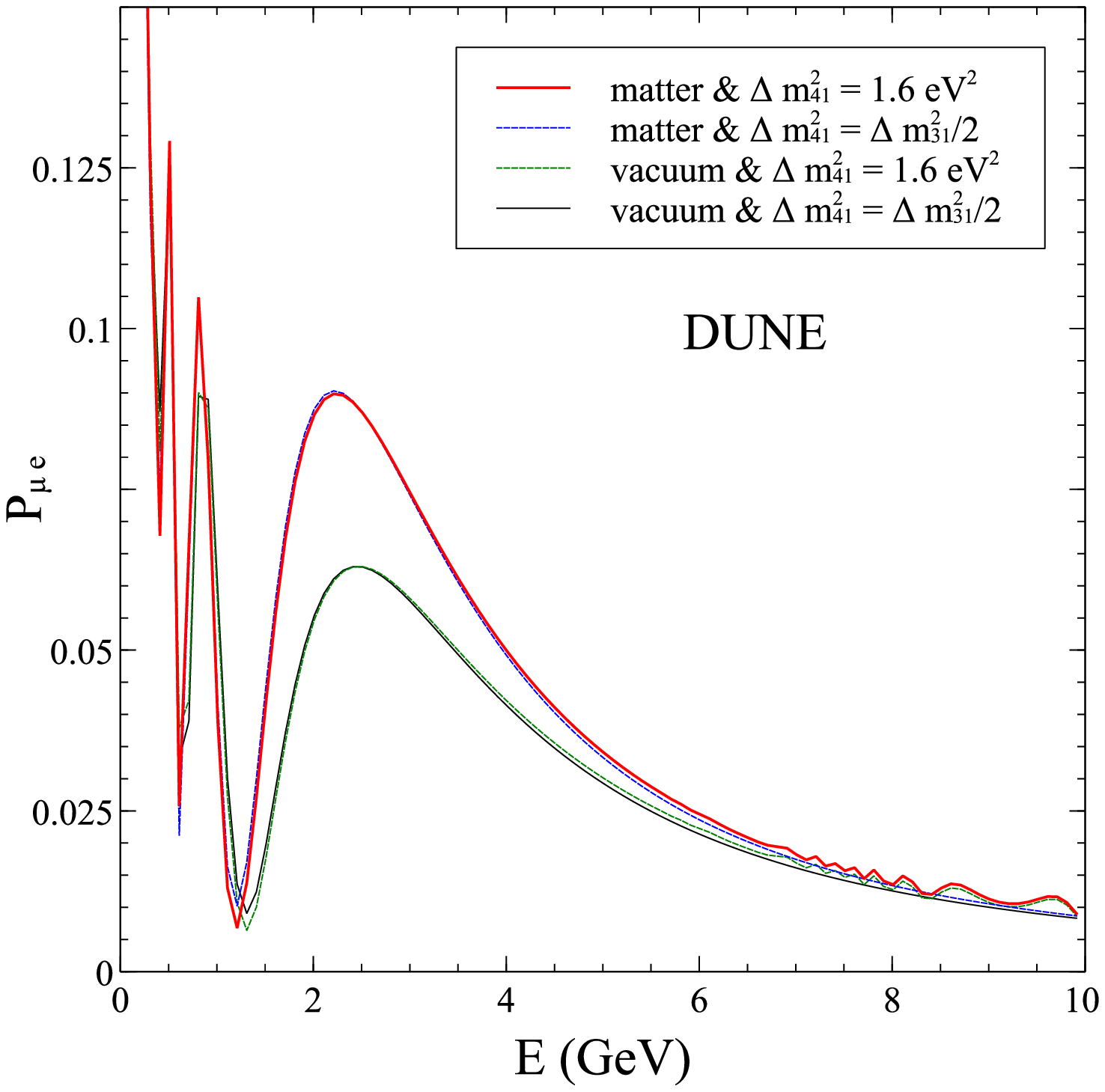}
\caption{\label{prob}
Appearance probability as a function of energy for T2HK (left panels) and DUNE (right panels). The solid curves are for the case $\Delta m^{2}_{41} = 1.6 $eV$^{2}$ after averaging over the fast oscillations induced by the high mass-squared difference, while the dashed curves are for the case when $\Delta m^{2}_{41} = \frac{1}{2}\ma$. The red solid and blue curves are for $\ms=7.5\times 10^{-3}$ eV$^2$ while the green and black curves are for the approximate case of $\ms=0$. The fast oscillations due $\Delta m^2_{41}$ has been averaged out by Gaussian smearing.}
\end{figure}

We will consider one additional sterile neutrino, a scenario referred to 3+1 \citep{Goswami:1995yq} in the literature. Here we have a 4$\times$4 mixing matrix described by 6 mixing angles: $\theta_{12}$, $\theta_{13}$, $\theta_{23}$, $\theta_{14}$, $\theta_{24}$ and $\theta_{34}$ and 3 phases: $\delta_{13}$, $\delta_{24}$ and $\delta_{34}$. We use the parametrisation convention where the mixing matrix is given as:
\begin{equation}
U^{3+1} = O(\theta_{34},\delta_{34})O(\theta_{24},\delta_{24})R(\theta_{14})R(\theta_{23})O(\theta_{13},\delta_{13})R(\theta_{12}) \,,                                
\end{equation} 
where $O(\theta_{ij},\delta_{ij})$ are 4$\times$4 orthogonal matrices with phase $\delta_{ij}$ associated with the $ij$ sector, and $R(\theta_{ij})$ are the rotation matrix with the $ij$ sector. In the 3+1 scenario, in addition to $\ms$ and $\ma$, there is a third mass-squared difference, $\Delta m^{2}_{41}$.  For the long-baseline experiments, fast oscillations due this third mass squared difference would be averaged out for values of $\Delta m^{2}_{41} \sim \mathcal{O}(1$eV$^{2}$) due to the finite resolution of the detectors. Hence, for these values of $\Delta m^{2}_{41} \sim$1 eV$^{2}$ the $\nu_\mu \to \nu_e$ oscillation probability can be written as \citep{Choubey:2017ppj}, 
\begin{equation}
P_{\mu e}^{4\nu} = P_{1} + P_{2}(\da)+P_{3}(\db)+P_{4}(\da+\db)
\,,
\label{eq:pmue_h}
\end{equation}
where $P_{1}$ is the term independent of any phase, $P_{2}(\da)$ depends only on $\da$, $P_{3}(\db)$ depends only on $\db$ and $P_{4}(\da+\db)$ depends on the combination ($\da+\db$). Note that while we will numerically calculate the exact probability in presence of earth matter effect in all our results presented in the next section, here for simplicity we give the analytic expressions for vacuum only. We stress that the same expressions will be valid for neutrinos propagating in constant density matter, just by replacing the oscillation parameters $\theta_{13}$ and $\ma$ in vacuum with those in constant density matter. We will discuss towards the end of this section the role of matter in the fake solution. 
The full expression of the different terms in Eq.~(\ref{eq:pmue_h}) are as follows:
\beqa
\label{eq:p1}
P_1 &=& \frac{1}{2}\sin^22\tmef
\nonumber \\
&&+ (X^2\sin^22\tmet - \frac{1}{4}\sin^22\ty\sin^22\tmef)
(\cos^{2}\theta_{12}\sin^{2}\Delta_{31}+\sin^{2}\theta_{12}\sin^{2}\Delta_{32})
\nonumber \\ &&+ (X^2Y^2-\frac{1}{4}X^2\sin^22\tx\sin^22\tmet
- \frac{1}{4}
\cos^4\ty\sin^22\tx\sin^22\tmef) \sin^2\Delta_{21}\,,\\
P_{2}(\delta_{13}) &=& YX^2  \sin2\tmet 
\big[\cos(\da)\big( \cos2\tx\sin^2\Delta_{21}+\sin^2\Delta_{31}-\sin^2\Delta_{32}\big)\nonumber \\ &&- \frac{1}{2} \sin(\da)
\big(\sin2\Delta_{21}-\sin2\Delta_{31}+\sin2\Delta_{32} \big)\big]\,,\\
\label{eq:p2}
P_{3}(\delta_{24})  &=&   XY \sin2\tmef 
\big[\cos(\db)\big(\cos2\tx\cos^2\ty\sin^2\Delta_{21}-\sin^2\ty(\sin^2\Delta_{31}
-\sin^2\Delta_{32}\big) \big)
\nonumber \\ &&+ \frac{1}{2} \sin(\db) 
\big(\cos^2\ty\sin2\Delta_{21}+\sin^2\ty(\sin2\Delta_{31}-\sin2\Delta_{32})
\big)\big]\,,\\
\label{eq:p3}
P_{4} (\delta_{13} + \delta_{24})  &=&  X \sin2\tmet \sin2\tmef
\big[\cos(\da + \db) \big(-\frac{1}{2}\sin^22\tx\cos^2\ty\sin^2\Delta_{21} 
\nonumber \\ &&+ \cos2\ty(\cos^2\tx\sin^2\Delta_{31}+\sin^2\tx\sin^2\Delta_{32})\big)
\nonumber \\ &&+ \frac{1}{2} \sin(\da + \db) 
\big(\cos^2\tx\sin2\Delta_{31}+\sin^2\tx\sin2\Delta_{32} \big)\big]\,,
\label{eq:p4}
\eeqa
where,
\begin{eqnarray}
\sin2\theta_{\mu e}^{3\nu}&=&\sin2\ty\sin\tz\\
\sin2\theta_{\mu e}^{4\nu}&=&\sin2\theta_{14}\sin\theta_{24}\\
X&=&\cos\theta_{14}\cos\theta_{24}\\
Y&=&\cos\theta_{13}\cos\theta_{23}\sin2\theta_{12}
\,,
\end{eqnarray}
 and,
 \beqa
  \Delta_{ij}=\frac{\Delta m^{2}_{ij}L}{4E}
  \eeqa
  If we put the approximation $\ms = 0$ and  the condition $\delta_{13}+\delta_{24} = 0$ in Eq. (\ref{eq:p1}-\ref{eq:p4}), we get, 
 \begin{eqnarray}\label{eq:pmueh_d0}
\nonumber P_{\mu e} &=& \frac{1}{2}\sin^{2}(2\theta^{4\nu}_{\mu e})\\
\nonumber &&+(X^{2}\sin^{2}(2\theta^{3\nu}_{\mu e})-\frac{1}{4}\sin^{2}(2\theta_{13})\sin^{2}(2\theta^{4\nu}_{\mu e}))\sin^{2}(\Delta_{31})\\
 &&+ X\sin(2\theta^{3\nu}_{\mu e})\sin(2\theta^{4\nu}_{\mu e})\cos(2\theta_{13})\sin^{2}(\Delta_{31})
\,.
\end{eqnarray} 
On the other hand, for values of $\Delta m^{2}_{41} << $1 eV$^{2}$, the oscillations due to this mass scale will survive the detector resolutions and show-up at the long-baseline detector. If we continue using the approximation $\ms = 0$ while allowing the $\Delta m^{2}_{41}$-driven oscillatory terms, the expression for the probability becomes, 
\begin{eqnarray}\label{eq:pmue}
\nonumber P_{\mu e} &=& \sin^{2}(2\theta^{4\nu}_{\mu e})[\sin^{2}(\theta_{13})\sin^{2}(\Delta_{43})+\cos^{2}(\theta_{13})\sin^{2}(\Delta_{41})]\\
\nonumber  &&-\cos(\delta_{13}+\delta_{24})X\sin(2\theta^{4\nu}_{\mu e})\sin(2\theta^{3\nu}_{\mu e}))[\sin^{2}(\Delta_{43})-\sin^{2}(\Delta_{41})]\\
\nonumber &&+ \frac{1}{2}\sin(\delta_{13}+\delta_{24})X\sin(2\theta^{4\nu}_{\mu e})\sin(2\theta^{3\nu}_{\mu e}))[\sin(2\Delta_{43})-\sin(2\Delta_{41})]\\
\nonumber &&+(X^{2}\sin^{2}(2\theta^{3\nu}_{\mu e})-\frac{1}{4}\sin^{2}(2\theta_{13})\sin^{2}(2\theta^{4\nu}_{\mu e}))\sin^{2}(\Delta_{31})\\
\nonumber &&+ \cos(\delta_{13}+\delta_{24})X\sin(2\theta^{3\nu}_{\mu e})\sin(2\theta^{4\nu}_{\mu e})\cos(2\theta_{13})\sin^{2}(\Delta_{31})\\
&&+ \frac{1}{2}\sin(\delta_{13}+\delta_{24})X\sin(2\theta^{3\nu}_{\mu e})\sin(2\theta^{4\nu}_{\mu e})\sin(2\Delta_{31})
\,.
\end{eqnarray}
If we use the condition $\delta_{24}+\delta_{13} = 0$ in Eq.~\eqref{eq:pmue} we get,
\begin{eqnarray}\label{eq:pmue_d0}
\nonumber P_{\mu e} &=& \sin^{2}(2\theta^{4\nu}_{\mu e})[\sin^{2}(\theta_{13})\sin^{2}(\Delta_{43})+\cos^{2}(\theta_{13})\sin^{2}(\Delta_{41})]\\
\nonumber  &&-X\sin(2\theta^{4\nu}_{\mu e})\sin(2\theta^{3\nu}_{\mu e}))[\sin^{2}(\Delta_{43})-\sin^{2}(\Delta_{41})]\\
\nonumber &&+(X^{2}\sin^{2}(2\theta^{3\nu}_{\mu e})-\frac{1}{4}\sin^{2}(2\theta_{13})\sin^{2}(2\theta^{4\nu}_{\mu e}))\sin^{2}(\Delta_{31})\\
 &&+ X\sin(2\theta^{3\nu}_{\mu e})\sin(2\theta^{4\nu}_{\mu e})\cos(2\theta_{13})\sin^{2}(\Delta_{31})
 \,.
\end{eqnarray}
Now, for the particular choice $\Delta m^{2}_{41} = \frac{1}{2} \ma$ the oscillation probability reduces to,
\begin{eqnarray}\label{eq:pmue_hlf}
\nonumber P_{\mu e} &=& \sin^{2}(2\theta^{4\nu}_{\mu e})\sin^{2}(\frac{\Delta_{31}}{2})\\
\nonumber &&+(X^{2}\sin^{2}(2\theta^{3\nu}_{\mu e})-\frac{1}{4}\sin^{2}(2\theta_{13})\sin^{2}(2\theta^{4\nu}_{\mu e}))\sin^{2}(\Delta_{31})\\
 &&+ X\sin(2\theta^{3\nu}_{\mu e})\sin(2\theta^{4\nu}_{\mu e})\cos(2\theta_{13})\sin^{2}(\Delta_{31})
 \,.
\end{eqnarray}
We can see from Eq.~(\ref{eq:pmueh_d0}) and Eq.~(\ref{eq:pmue_hlf}) that the oscillation probability for these two cases are equal, except for the first term. But at the oscillation maximum, $\Delta_{31} = \frac{\pi}{2}$,  the first term of the Eq. \eqref{eq:pmue_hlf} becomes $\frac{1}{2}\sin^{2}2\theta^{4\nu}_{\mu e}$. So at the oscillation maximum the oscillation probability for the $\Delta m^{2}_{41} > $1 eV$^{2}$ case given by Eq.~(\ref{eq:pmueh_d0}) becomes  exactly equal to the oscillation probability for the $\Delta m^{2}_{41} = \frac{1}{2} \ma$ case given by Eq.~(\ref{eq:pmue_hlf}). This produces a fake solution in $\Delta m_{41}^2$ at the energy corresponding to $\Delta m_{31}^2$ oscillation maximum for a given baseline $L$. Note that the above condition for the fake solution was obtained for oscillations in vacuum. As mentioned before, using the corresponding expression in constant density matter we would get the fake solution in matter at the energy that gives the oscillation maximum for ${\Delta m_{31}^2}^m$ for a given baseline $L$. One can check that for earth matter effect in experiments like T2HK and DUNE the energy at which we get oscillation maximum for  ${\Delta m_{31}^2}^m$ is very similar to that for ${\Delta m_{31}^2}$. We will quantify the effect of matter on the fake solution shortly in Fig.~\ref{prob}.

For energies away from the ${\Delta m_{31}^2}$-driven first oscillation maximum, the oscillatory term $\sin^{2}(\frac{\Delta_{31}}{2} )\neq 1/2$ exactly and hence, the above mentioned fake solution is not exact. However, we see that the fake solution is achieved even outside the $\Delta m_{31}^2$-driven first oscillation maximum, albeit approximately. To explore this further, let $P^{\ma/2}$ be the oscillation probability for $\Delta m^{2}_{41} = \ma/2 $ and $P^{ave}$ for $\Delta m^{2}_{41} \sim 1 $eV$^{2}$, and let $\Delta P$ be the difference between $P^{\ma/2}$ and $P^{ave}$:
\be
\Delta P &=& P^{\ma/2}-P^{ave} \,,
\nonumber \\
 &=& \sin^{2}(2\theta^{4\nu}_{\mu e})\bigg[\sin^{2}\bigg(\frac{\ma L}{8E}\bigg)-\frac{1}{2}\bigg]
 \,.
\ee
The term $\sin^{2}\Big(\frac{\ma L}{8E}\Big)$  takes values only in the range [0,1]. Hence, $\Delta P$ is also bounded, and lies between $-0.5\sin^{2}(2\theta^{4\nu}_{\mu e})\leq \Delta P \leq 0.5\sin^{2}(2\theta^{4\nu}_{\mu e})$. Therefore, for $\sin^{2}(2\theta^{4\nu}_{\mu e})$ small, $\Delta P_{max}$ is also expected to be small. For example, if  $\sin^{2}(2\theta^{4\nu}_{\mu e})\sim 10^{-3}$, then $\Delta P_{max} \sim \mathcal{O}(10^{-3})$, whereas the oscillation probability $P_{\mu e} \sim \mathcal{O}(10^{-2})$, so we can say that there is an approximate degeneracy for all energies for small values of $\sin^{2}(2\theta^{4\nu}_{\mu e})$. In our simulations we have used the value of $\sin^{2}(2\theta^{4\nu}_{\mu e})$ that lies within the currently allowed range obtained from global analysis of all short baseline data including LSND and MiniBooNE \citep{Gariazzo:2017fdh} and corresponds to $10^{-3}$ in our case.


The Fig.~\ref{prob} shows the probability curves for T2HK (left panels) and DUNE (right panels) as a function of energy. The solid curves are for the case $\Delta m^{2}_{41}=1.6$ eV$^{2}$ after averaging over the fast oscillations induced by the high mass-squared difference, while the dashed curves are for the case when $\Delta m^{2}_{41} = \frac{1}{2}\ma$. In the top panels we show the impact of non-zero $\ms$ on the fake solution while in the bottom panels we show the effect of matter on the fake solution. In the top (bottom) panels the red solid and blue curves are for $\ms=7.5\times 10^{-3}$ eV$^2$ (matter) while the green and black curves are for the case of $\ms=0$ (vacuum). All curves are drawn for $\Delta m_{31}^2=2.75\times 10^{-3}$ eV$^2$, $\delta_{13}+\delta_{24}=0$, $\theta_{13}=8.5^\circ$, $\theta_{23}=45^\circ$, $\theta_{14}=8.13^\circ$, $\theta_{24}=7.14^\circ$ and $\theta_{34}=0$. From Fig.~\ref{prob}, the existence of the fake solution between high values of $\Delta m^{2}_{41}$ and $\Delta m^{2}_{41}=\frac{1}{2}\ma$ can be easily seen. The probabilities for the two cases mentioned above are almost same. Very small differences are seen at energies away from the oscillation maximum, for reasons discussed above. Due to the finite resolution of the detector, this tiny difference is not expected to be differentiated by the detector and hence we expect to see two degenerate solutions for two values of $\Delta m^{2}_{41}$. 


A final comment is in order. Fake solutions in the active-sterile sector have been studied in the context of Daya Bay \cite{An:2016luf}, RENO and Double Chooz \cite{Bandyopadhyay:2007rj} experiments. The Daya Bay paper \cite{An:2016luf} discusses the degeneracy between $\sin^2\theta_{13}$ and $\sin^2\theta_{14}$ for the case when $\Delta m_{31}^2 \approx \Delta m_{41}^2$. In \cite{Bandyopadhyay:2007rj} the degeneracy between $\sin^2\theta_{13}$ and $\sin^2\theta_{14}$  appears in the $P_{\bar e\bar e}$ channel in the 3+2 scenario when $\Delta m_{41}^2$ and $\Delta m_{51}^2$ is taken to be in the 1 eV$^2$ regime. In this paper we will look into the degeneracy in the $P_{e\mu}$ channel in the context of accelerator-based long-baseline experiments. The muon neutrino disappearance is seen to not have any active-sterile degeneracy. 

\section{Numerical results \label{sec:ext}}


We have used GLoBES \citep{Huber:2004ka,Huber:2007ji} and the necessary code \cite{Kopp:2006wp,Kopp:2007ne} for sterile neutrino oscillation for the simulation of the long-baseline experiments. For this work we have taken baseline of 295 km and fiducial volume of 374 kton with 1.3 MW beam, 2.5$\degree$ off-axis from the beam axis for T2HK and 1300 km baseline with 34 kton  liquid Argon detector for DUNE. The energy resolution is taken as 15\%$/\sqrt{E}$. For all our analysis, we used $\ms = 7.5\times10^{-5}$eV$^2$,  $\ma = 2.5 \times 10^{-3}$ eV$^{2}$, $\theta_{12} = 33.48\degree$, $\theta_{13} = 8.5\degree$, $\theta_{23}. = 45\degree$ and $\delta_{13} = -90\degree$ for the standard mixing parameters, consistent with the current best-fit \citep{Esteban:2018azc}. And for sterile sector, $\theta_{14} = 8.13 \degree $, $\theta_{24} = 7.14\degree$ are taken and $\theta_{34}$ and $\delta_{34}$ are taken to be zero as they have very small effect. The phase $\delta_{24}$ and $\Delta m^{2}_{41}$ will be mentioned for different plots shown in the results section. 

\begin{figure}
\includegraphics[width=0.45\textwidth]{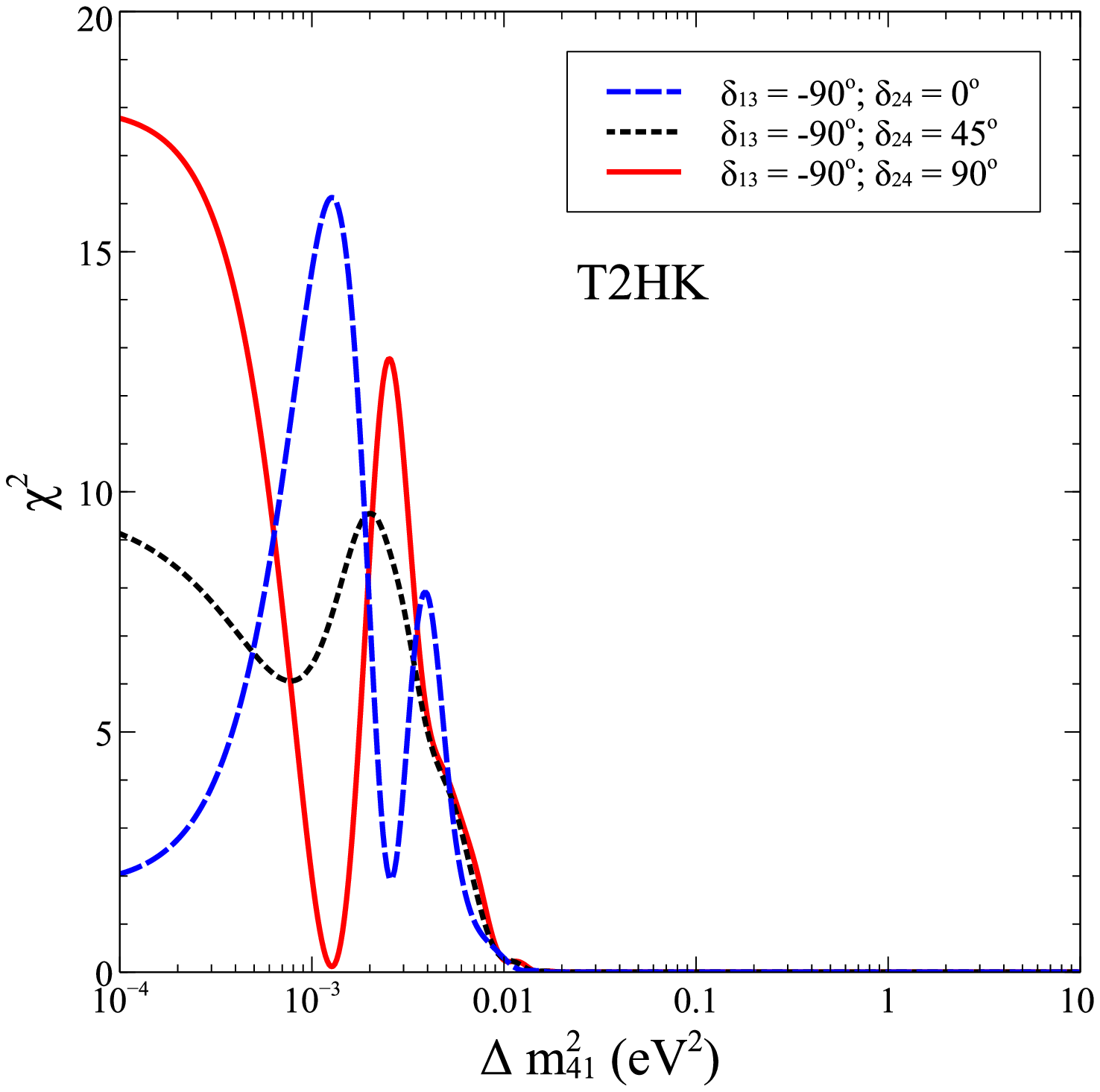}
\includegraphics[width=0.45\textwidth]{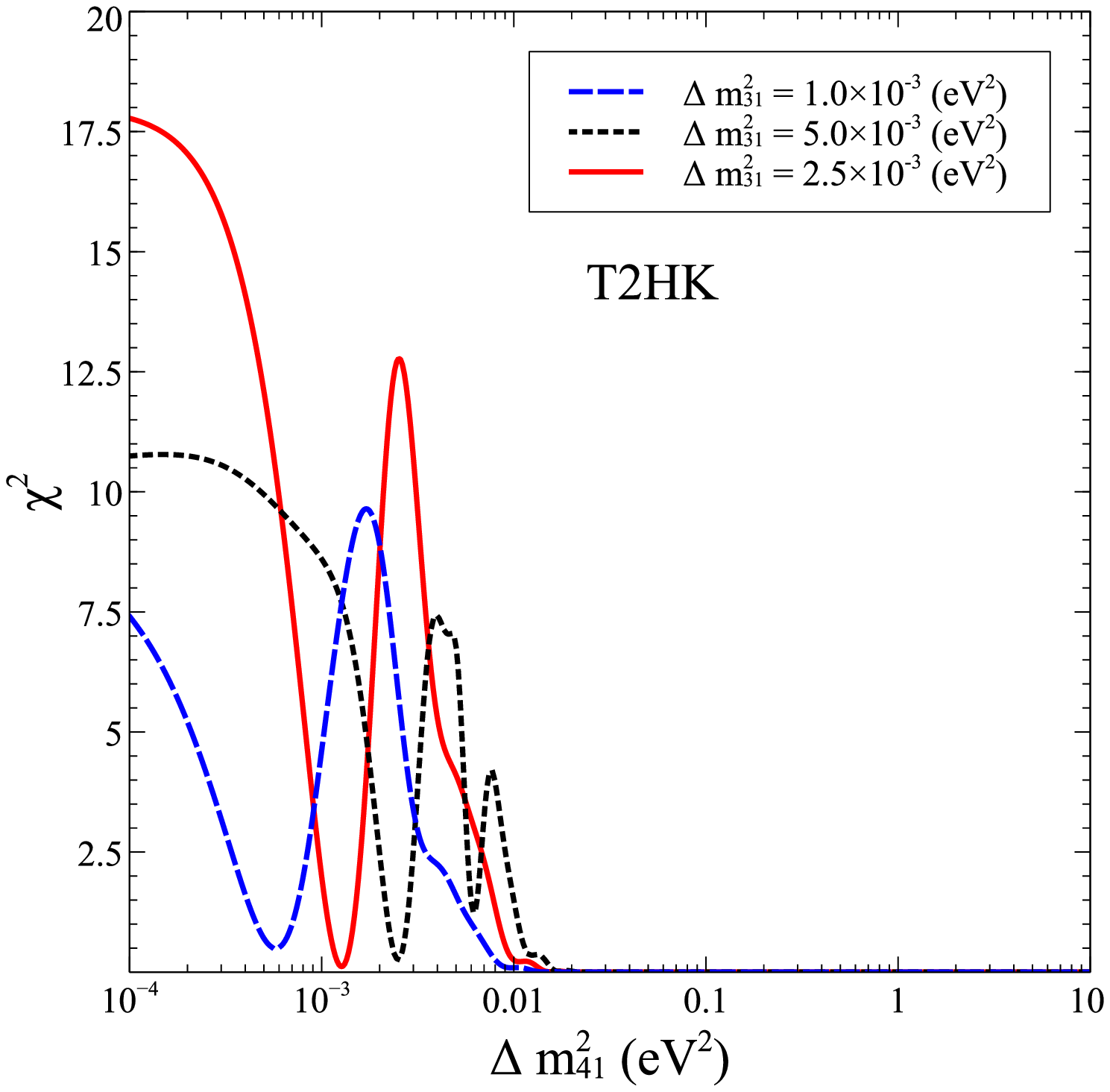}
\includegraphics[width=0.45\textwidth]{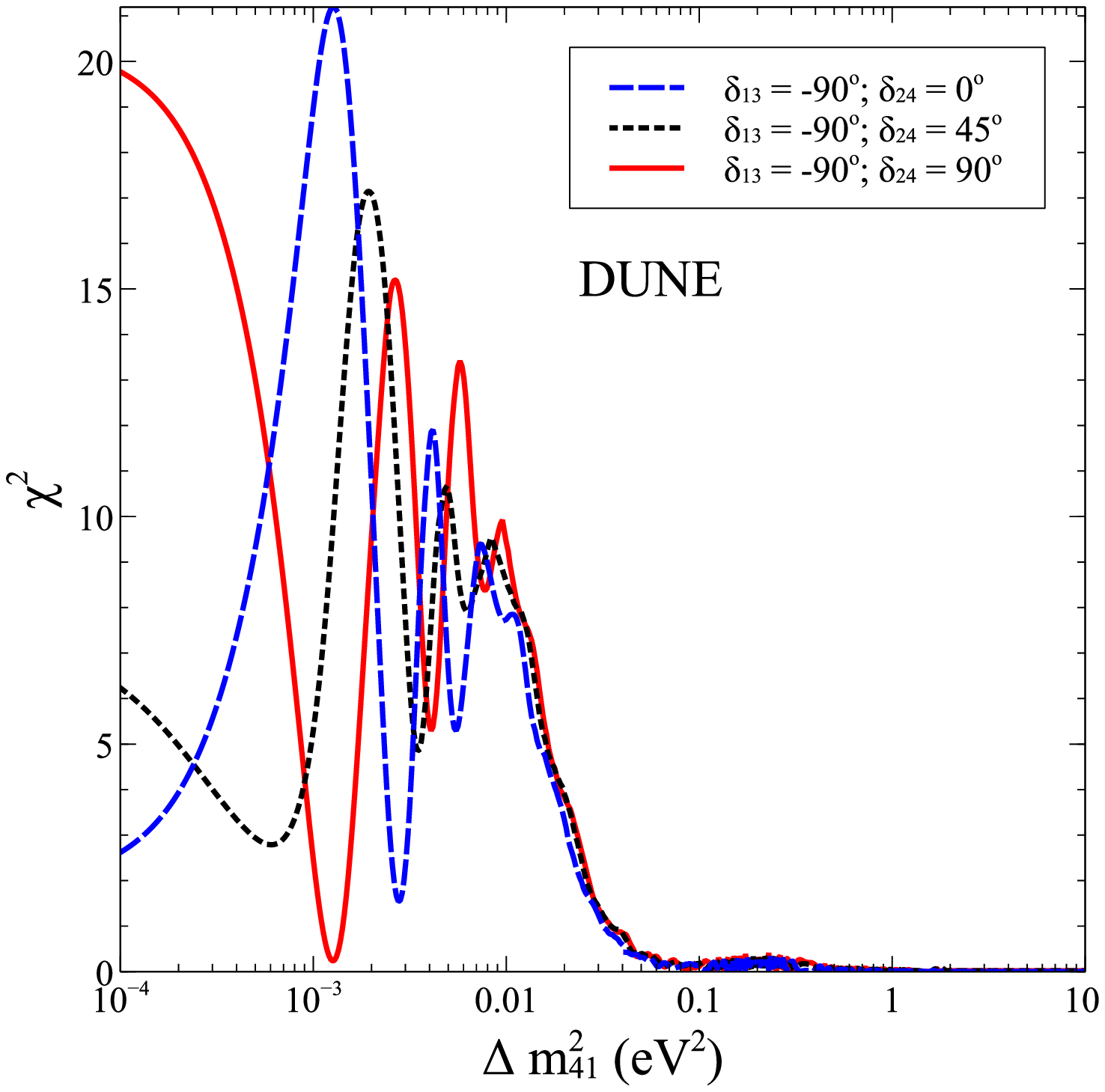}
\includegraphics[width=0.45\textwidth]{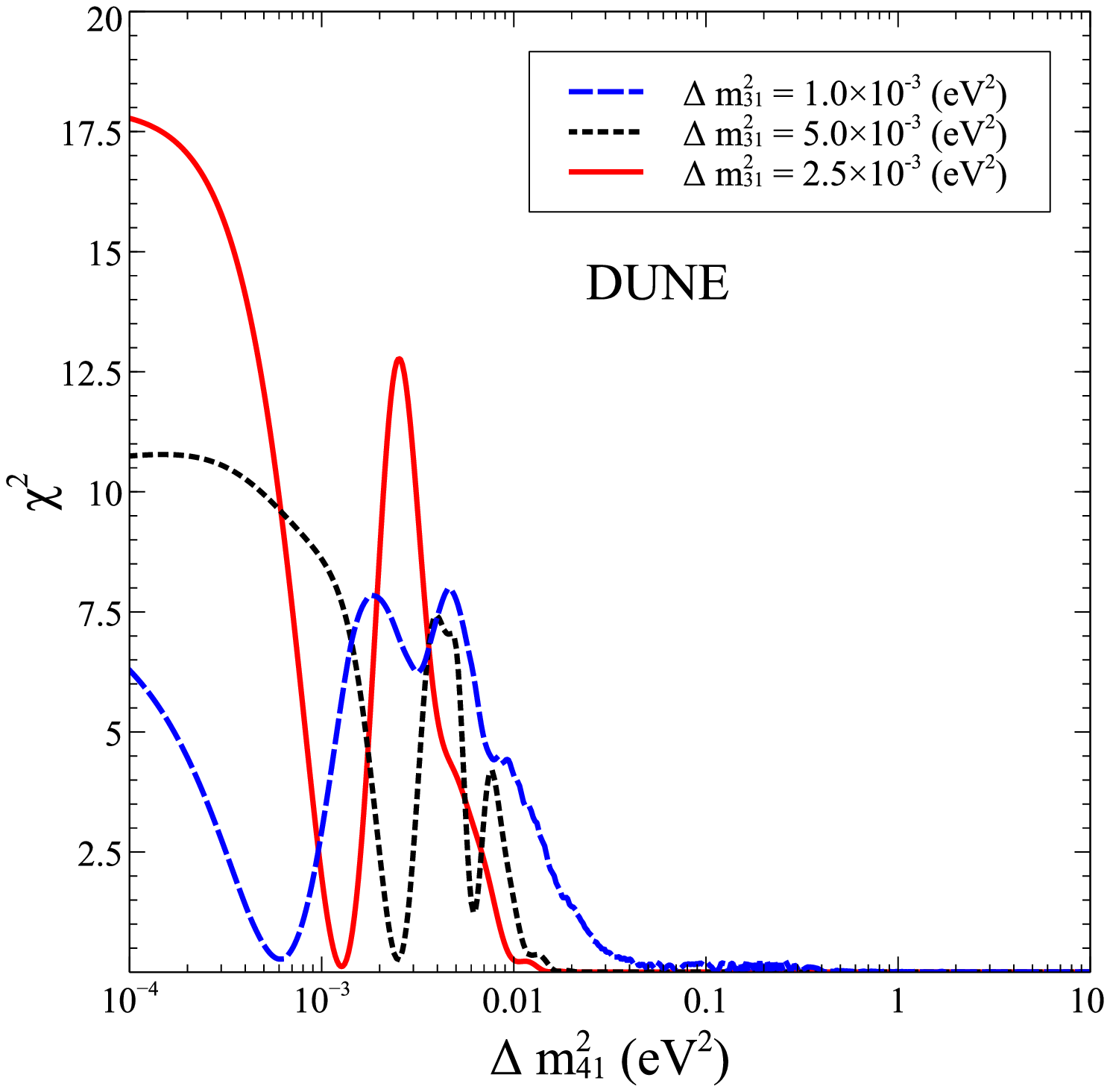}
\caption{\label{chi_del41_del24} 
$\chi^{2}$ as a function of $\Delta m^{2}_{41}$ for different combinations of CP-violating phases (left panels) and various values of $\ma$ (right panels). Top panels are for T2HK while bottom panels are for DUNE. }
\end{figure}

The Fig.~\ref{chi_del41_del24} shows the $\chi^{2}$ as a function of $\Delta m^{2}_{41}$ (test) for appearance channel only. The top panels are for T2HK while the bottom panels are for DUNE.  In the left panels we show different $\chi^2$ curves for different combinations of CP-violating phases $\delta_{13}$ and $\delta_{24}$ while in the right panels we show different curves for different values of $\ma$ in the data. In all panels the data have been generated for $\Delta m^{2}_{41} = 1.6$ eV$^{2}$.  From the left panels we can see that when $\delta_{13}= -90^{\circ}$ and $\delta_{24} = 90^{\circ}$, we get two (nearly degenerate) solutions in $\Delta m_{41}^2$. One solution corresponds to all values $\Delta m_{41}^2 \gtap 0.01$ eV$^2$ for which $\Delta m_{41}^2$-driven oscillations are averaged out at the far detector of the experiment. This is the true solution since for all these values of $\Delta m_{41}^2$ the fast oscillations are averaged out as for $\Delta m_{41}^2=1.6$ eV$^2$ in the data. In addition, there is a fake solution at $\Delta m_{41}^2\simeq (1/2)\Delta m_{31}^2$ at which the value of $\chi^{2}$ is nearly zero. This degeneracy corresponding to $(\delta_{24}+\delta_{13}) = 0$ was explained in the previous section \ref{sec:mixangle} in terms of the probabilities, both analytically as well as numerically. The fact that the $\chi^2$ at the fake solution is slightly different from zero comes mainly due to two reasons - (i) in this plot we have taken $\Delta m_{21}^2$ non-zero, and (ii) as discussed before, the $\chi^2$s have been obtained using the full energy range of T2HK and DUNE, while the exact degeneracy exists only at the oscillation maximum. 

Note from the figure that for all other combinations of $\delta_{13}+\delta_{24}$, the $\chi^2$ at $\Delta m_{41}^2\simeq (1/2)\Delta m_{31}^2$ is far from zero. However, for both the other combinations of $\delta_{13}+\delta_{24}$ plotted in the figure, there are dips in the value of $\chi^2$ at values of $\Delta m_{41}^2$ other than the true values corresponding to the data. Therefore, one can expect to see islands of allowed $\Delta m_{41}^2$ at values close to $(1/2)\Delta m_{31}^2$ for all combinations of $\delta_{13}+\delta_{24}$. While for $\delta_{13}+\delta_{24}=0$, this fake island is expected to appear as a nearly degenerate solution, for all other combinations it would appear only at a higher C.L. Note also that though for T2HK there is dip at only one fake value of $\Delta m_{41}^2$, for DUNE there are multiple dips. This is because DUNE is a wide-band beam experiment while T2HK is a narrow-beam experiment. 

The fake solution is further explored in the right panels of Fig.~\ref{chi_del41_del24}. Here the data is generated for only the case $\delta_{13}+\delta_{24}=0$ with $\delta_{13} = -90\degree $ and $\delta_{24} = 90\degree$ but for different values of $\ma$. The red solid, blue dashed and the black dotted curves are for $\ma = 2.5\times10^{-3}$ eV$^{2}$, $\ma = 1.0\times10^{-3}$ eV$^{2}$ and $\ma = 5.0\times10^{-3}$ eV$^{2}$, respectively. As the $\delta_{13}+\delta_{24} = 0$, there are degeneracies present in all of the three curves. And it is interesting to note that the fake solution is always at half of the value $\ma$.  

\begin{figure}
\includegraphics[width=0.45\textwidth]{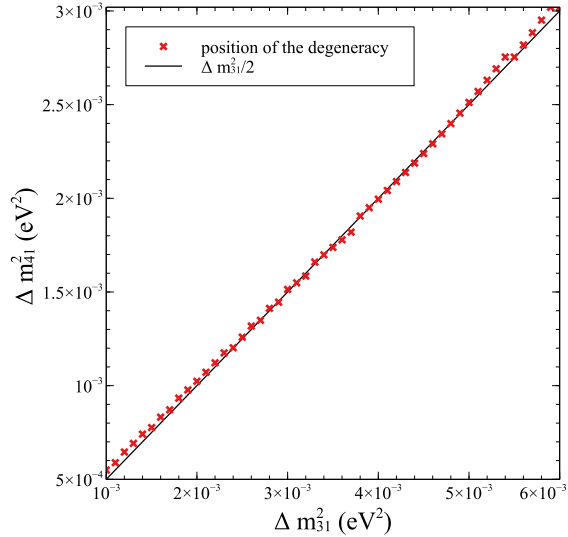}
\caption{\label{delm41}
The figure shows the positions of the fake solutions. The red data points show the positions of the fake solutions in $\Delta m^{2}_{41}$ as a function of $\ma$ and the black straight line is a linear fit to red points corresponding to $\Delta m_{41}^2=(1/2)\ma$. }
\end{figure}

In Fig.~\ref{chi_del41_del24} we saw that the fake solution of $\Delta m_{41}^2$ always appears at half of $\ma$. In Fig.~\ref{delm41} we further illustrate it for  T2HK. The plot for DUNE is identical and not shown here for brevity. The x-axis represents the value of $\ma$ in the data and the red dots represent the corresponding value of the fake $\Delta m^{2}_{41}$ solution when the true solution is taken at $\Delta m_{41}^2=1.6$ eV$^2$. The black line corresponds to $\Delta m_{41}^2 = \frac{1}{2}\ma$. We can see from Fig.~\ref{delm41} that this line fits the red points very well, so indeed the fake solution is at half of the value $\ma$. 

\begin{figure}
\includegraphics[width=0.45\textwidth]{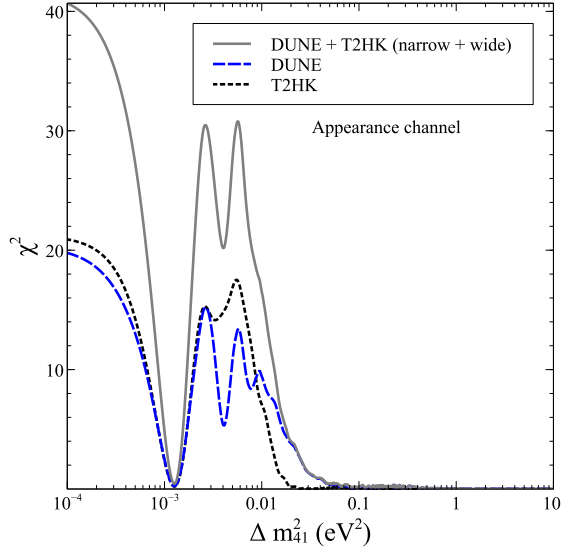}
\caption{\label{narrow_wide} $\chi^{2}$ as a function of $\Delta m^{2}_{41}$ for different types of experiments. The blue dashed curve represents DUNE (wide band beam), the black dotted curve represents T2HK, (narrow band beam), and the grey solid curve shows their combination.}
\end{figure}

 The Fig.~\ref{narrow_wide} shows the effect of combining the data from a  narrow band beam experiment (T2HK) with that from a wide band beam experiment (DUNE). The black dotted curve shows the $\chi^{2}$ for the narrow band experiment, the blue dashed curve shows the $\chi^{2}$ for the wide band experiment, and the grey solid curve shows the combination of them. We can see from Fig.~\ref{narrow_wide} that even though the grey curve has a higher $\chi^2$ than the blue and black curves for some values of $\Delta m_{41}^2$, but the fake solution is still present. 
This is expected because the fake solution is seen to be present in both the narrow as well as the wide band beam experiments at the same value of $\Delta m^{2}_{41}$. Therefore, the combination of these two types of experiments can not remove the fake solution. Note however that while the fake solution survives, its range is reduced if we combine the narrow and wide band beam experiments because of increased total statistics. 

\begin{figure}
\includegraphics[width=0.45\textwidth]{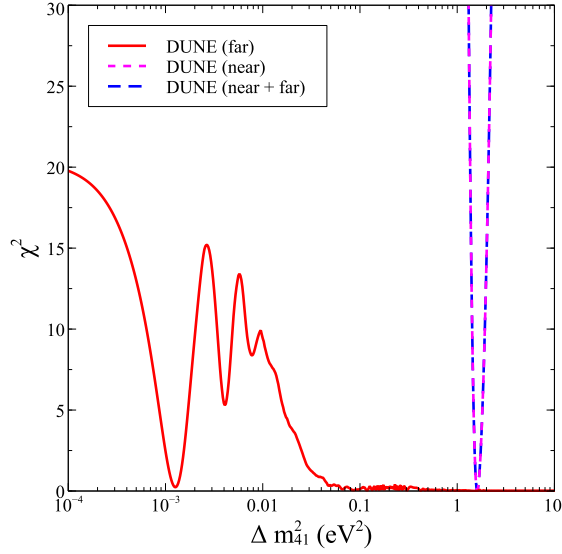}
\caption{\label{near_far} $\chi^{2}$ as a function of $\Delta m^{2}_{41}$ for both near and far detectors. The red solid curve is for DUNE far detector only, the magenta dotted curve is for DUNE near detector, and the blue dashed curve is for near and far detector together.}
\end{figure}

The Fig.~\ref{near_far} shows the effect of the near detector. The red solid curve gives the $\chi^{2}$ for the DUNE far detector, the magenta dashed dotted curve is for the near detector and the blue dashed curve is for both near and far detector together. This figure shows the importance of the near detector. We can see from the figure that the fake solution completely goes away as soon as we add the near detector data in the analysis. This is expected since we do not expect the degeneracy in the detector. As a result, due to the much higher statistics in the near detector, the fake solution coming from the far detector data completely disappears.

\begin{figure}
\includegraphics[width=0.45\textwidth]{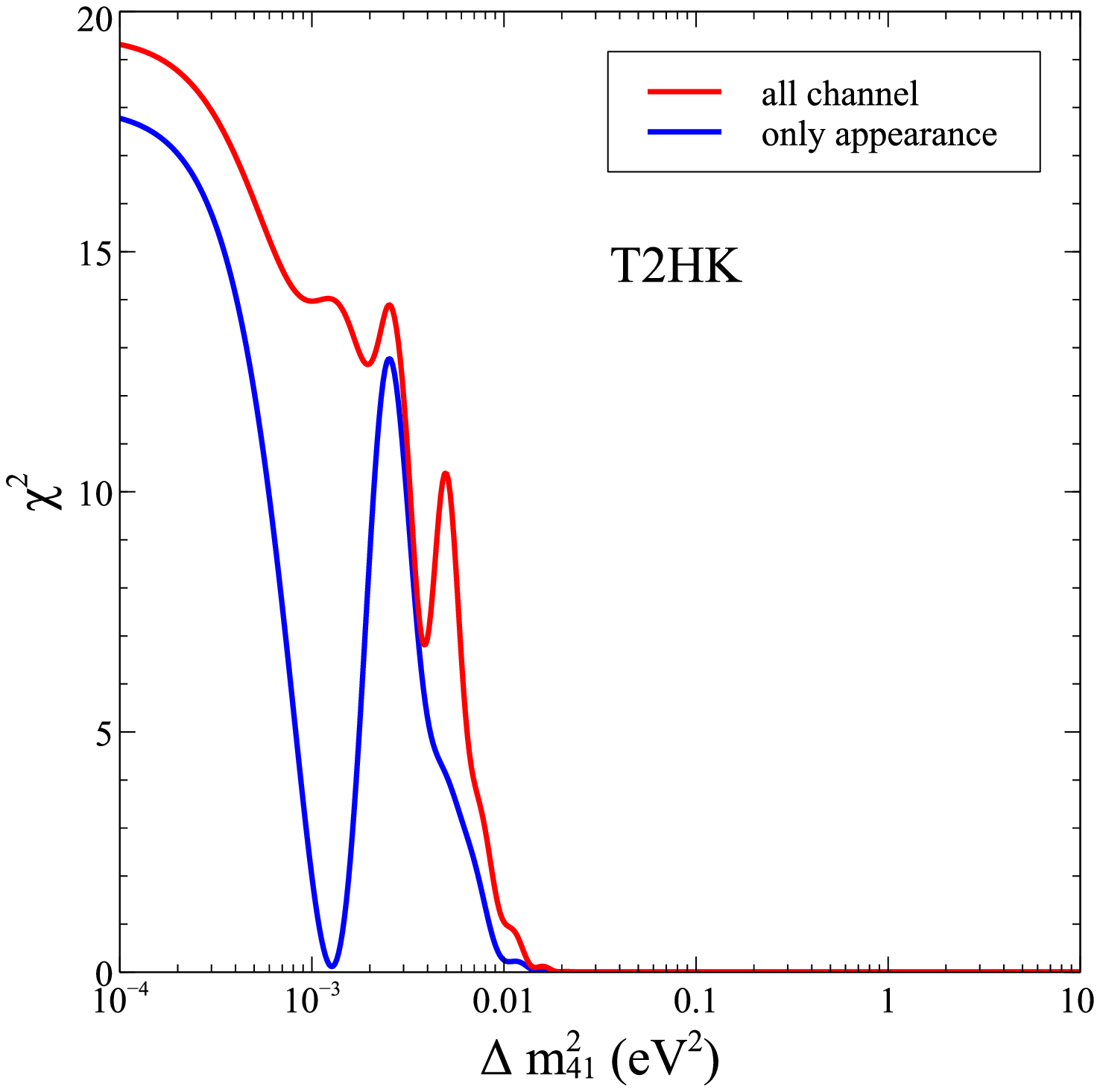}
\includegraphics[width=0.45\textwidth]{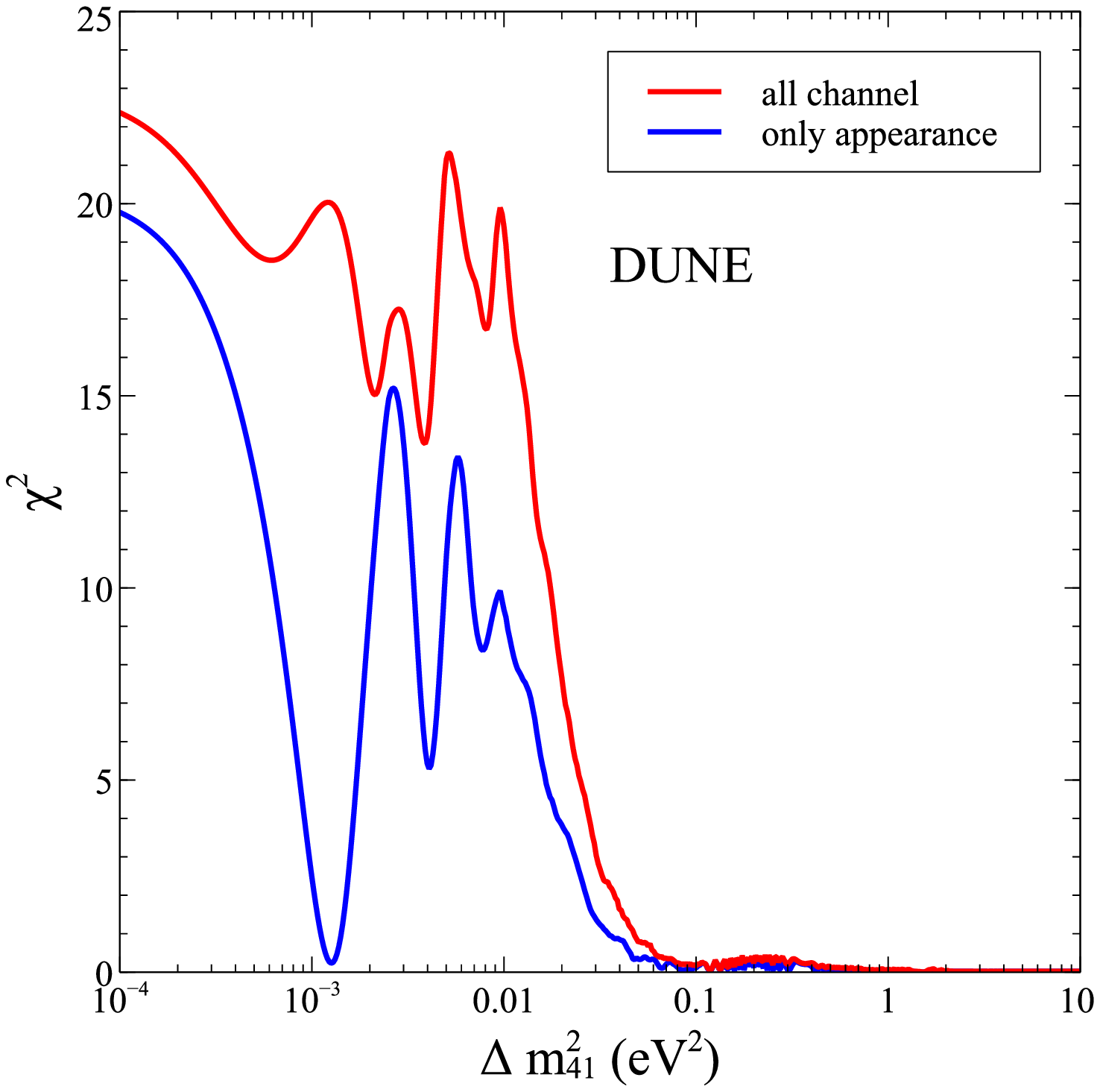}
\caption{\label{appear} Impact of disappearance channel on removing the $\Delta m_{41}^2$ fake solution. The blue lines show the $\chi^2$ as a function of $\Delta m_{41}^2$ (test) for the appearance channel only for T2HK (left panel) and DUNE (right panel). The red lines show the corresponding $\chi^2$ when both disappearance and appearance data are taken together in the fit.
}
\end{figure}

In all figures shown so-far, we have only considered the appearance channel in the long-baseline experiments. This is because the fake solution presented in this paper comes only in the appearance channel. The disappearance channel does not have this fake solution and hence should water down the emergence of the fake $\Delta m_{41}^2$ island. In Fig.~\ref{appear} we show the impact of adding the disappearance data into the analysis. The blue curve shows the $\chi^2$ as a function of $\Delta m_{41}^2$ for the appearance channel only, while the red curve shows the results when data from both appearance and disappearance channels are combined to obtain the $\chi^2$. We can see from the figure that introduction of disappearance channels removes the fake solution. The left panel is for T2HK and the right panel is for DUNE. The impact of the disappearance channel in lifting the degeneracy is visible in both experiments.

\section{Conclusions \label{sec:conclusion}}

Although the three neutrino oscillation paradigm is well established, there are tantalising hints for the possible existence of sterile neutrino state(s). While the question on whether sterile neutrinos driving such short-baseline neutrino flavor oscillations exist or not is far from settled, the presence of active-sterile mixing has been shown to impact neutrino oscillation at long-baseline experiments. Detailed analyses have been performed in the literature looking at the potential of long-baseline experiments in measuring the active-sterile oscillation parameters. In this paper we showed, for the first time, the existence of fake solution in the $P_{\mu e}$ appearance probability, coming from $\Delta m_{41}^2$. We showed that if the condition $\delta_{13}+\delta_{24}=0$ is satisfied,  fake solution is obtained in $P_{\mu e}$ at the oscillation maximum for $\Delta m^{2}_{41} = (1/2)\ma$ in addition to the true solution. We showed this both numerically as well as analytically. It was shown from a comparison of the numerical and analytical solutions that $\ms$ does not affect the position of the $\Delta m_{41}^2$ fake solution. We discussed the breaking of the degeneracy for values of energy away from the oscillation maximum and showed that this breaking is very mild. The impact of earth matter effects on the degeneracy was also discussed and shown to be a very mild shift in the energy of oscillation maximum.

We next showed the appearance of the degenerate $\Delta m_{41}^2$ solution from a $\chi^2$ analysis of the prospective data at both T2HK and DUNE. The degeneracy was shown to appear at $\Delta m^{2}_{41} = (1/2)\ma$, while other degenerate solutions in $\Delta m^{2}_{41}$ appeared at higher C.L. It was also seen that DUNE had multiple fake solutions at different $\Delta m^{2}_{41}$, in addition to the degenerate solution. Finally, we discussed that the degeneracy between large $\Delta m^{2}_{41} $ and $\Delta m^{2}_{41} = (1/2)\ma$ does not exist for the disappearance channel. As a result combining of the appearance and disappearance data resulted in erasing of the fake solution at $\Delta m^{2}_{41} = (1/2)\ma$. 

In conclusion, an fake solution in $\Delta m^{2}_{41}$ exists in the appearance probability at long-baseline experiments. This would result in one or more fake islands appearing in the $\Delta m^{2}_{41}$ allowed C.L. contours coming from the appearance channel of long-baseline experiments.  The disappearance channel does not have any such fake solution.

\section*{Acknowledgment}
We acknowledge the HRI cluster
computing facility (http://cluster.hri.res.in). 
This project has received funding from the European Union's Horizon
2020 research and innovation programme InvisiblesPlus RISE
under the Marie Sklodowska-Curie
grant  agreement  No  690575. This  project  has
received  funding  from  the  European
Union's Horizon  2020  research  and  innovation
programme  Elusives  ITN  under  the 
Marie  Sklodowska-Curie grant agreement No 674896.

\bibliography{ref}

\end{document}